\documentclass[aps,prd,twocolumn,groupedaddress,floatfix,showpacs,amsmath,amssymb]{revtex4}

\usepackage{graphicx}
\usepackage{bm}
\usepackage{dcolumn}

\begin{document}


\title{Changing universe model of redshift}


\author{John C. Hodge}
\email[]{jch9496@blueridge.edu}
\homepage[]{http://web.infoave.net/~scjh}
\altaffiliation{visiting from: XZD Corp., 3 Fairway St., Brevard, NC, 28712, scjh@citcom.net }
\affiliation{Blue Ridge Community College, 100 College Dr., Flat Rock, NC, 28731-1690}


\date{\today}

\begin{abstract}
The changing universe model (CUM) describes galaxy parameter relationships (SESAPS '03, session EB 2).  The CUM must be successfully applied to cosmological scale observations to be considered a cosmological model.  A major component of current cosmological models is the Hubble constant $H_\mathrm{o}$.  An equation is derived using the CUM model relating redshift $z$ and the distance $D$ to galaxies and is applied to a sample of 32 spiral galaxies with $D$ calculated using Cepheid variable stars.  The equation predicts a galaxy may have $z<0$ in special circumstances.  Three elliptical galaxies with peculiar characteristics are discovered to be CUM Sinks.  The Sinks give a physical explanation of the ``Virgocentric infall'' and ``Great Attractor'' observations without a large, unobserved mass.  At low cosmological distances, the equation reduces to $z \approx \exp(KD) \, -1 \approx KD$, where $K$ is a constant, positive value.  The equation predicts $z$ from galaxies over 23 Gpc distant approaches a constant value on the order of 1000.  The CUM gives a physical basis for the Doppler shift of particle photons.

\end{abstract}


\maketitle

\section{INTRODUCTION}
The changing universe model (CUM) \cite{hodg4} was derived from examining galaxy scale observations.  The CUM must be successfully applied to cosmological scale observations to be considered a cosmological model.  

That the redshift $z$ of emitted photons from galaxies generally increases with the extragalactic distance $D$ (Mpc) between the emitting galaxy and observer is well known ($z - D$ relationship).  The $z - D$ relationship has different meanings in current cosmologies.  The idea of ``expanding space'' may be defined as a set of comoving coordinates wherein Space is like an expanding rubber sheet that carries matter along with it.  Space is expanding between particles.  Alternatively, expanding Space may be like an expanding gas in a vacuum.  Space exerts a force on particles causing matter to be carried along with the expanding Space.  An alternate idea is that Space is an unobtrusive medium through which matter moves in free particle motion with a certain initial velocity and inertia.  In these cases, photons have a dual particle and wave nature and the distribution of matter in the universe is assumed isotropic and homogeneous in cosmological scale volumes.

Currently fashionable, cosmological models attributes $z$ to be solely a Doppler shift of a light wave.  The ``standard model'' assumes the universe is homogeneous and isotropic.  The assumption of homogeneity is compatible with either a static galaxy distribution or with a very special velocity field obeying the Hubble Law \cite[page~396]{binn}
\begin{equation}
D =  \frac{c}{H_\mathrm{o}}z
\label{eq:1},
\end{equation}
where $H_\mathrm{o}$ (km~s$^{-1}$~Mpc$^{-1}$) is the Hubble constant and $c$ (km~s$^{-1}$) is the speed of light.  The $H_\mathrm{o}$ occupies a pivotal role in current cosmologies.  The methods of calculating supernova distances, the cosmological microwave background (CMB) power spectrum, weak gravitational lensing, cluster counts, baryon oscillation, expansion of the universe, and the fundamental aspects of the Big Bang depend on $H_\mathrm{o}$ and the Hubble law.

\begin{figure}
\includegraphics[width=0.5\textwidth]{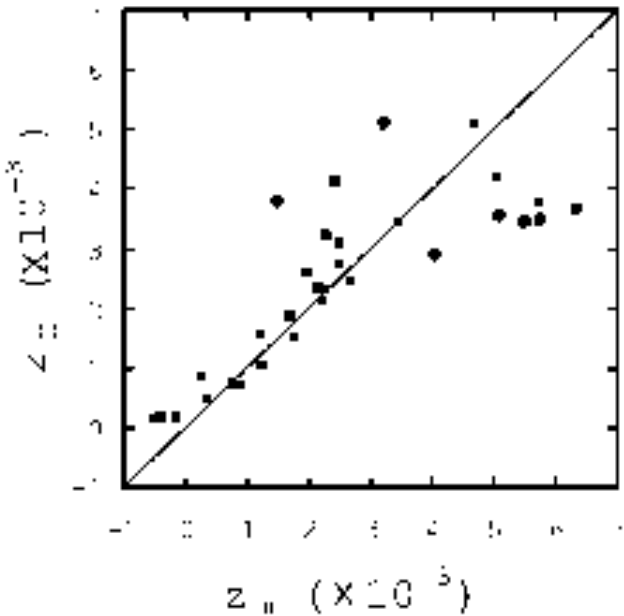}
\caption{\label{fig:1} Plot of the measured redshift $z_\mathrm{m}$ versus the calculated redshift $z_\mathrm{H}$ after the third iteration using Eq.~(\ref{eq:1}) and $D$ calculated using Cepheid variable stars for 32 galaxies \cite{free,macr}.  The straight line is a plot of $z_\mathrm{H} = z_\mathrm{m}$.  The circles indicate the data points for galaxies with (l,b) = (290$^\circ \pm 20^\circ$,75$^\circ \pm 15^\circ$).}
\end{figure}

However, the determination of $H_\mathrm{o}$ has a large uncertainty and different researchers calculate different values.  Figure~\ref{fig:1} shows the measured galactocentric redshift $z_\mathrm{m}$ versus the calculated redshift $z_\mathrm{H}$ using Eq.~(\ref{eq:1}), $H_\mathrm{o}=70$ km~s$^{-1}$~Mpc$^{-1}$, and $D$ calculated using Cepheid variable stars for 32 galaxies \cite{free,macr}.  The correlation coefficient of $z_\mathrm{H}$ versus $z_\mathrm{m}$ is 0.80.

The deviation of the recession velocity of a galaxy from the straight line of the Hubble Law is ascribed to a ``peculiar velocity'' of the photon emitting galaxy relative to earth \cite[page 439]{binn}.  The deviation from the Hubble Law is the only means of determining the peculiar velocity of a galaxy.  Therefore, to determine the distances to $n$ galaxies requires $n+1$ constants.  However, the peculiar velocity is assumed to be limited to less than 1000 km~s$^{-1}$, which is a small error at cosmological scale distances.  Also, the average peculiar velocity for all galaxies is assumed to be zero on the scale that the universe appears homogenous.  The 2dFGRS \cite{peac} suggests this scale is $z > 0.2$.

The circles in Fig.~\ref{fig:1} denote data for galaxies in the general direction of $( l, b)=(290^\circ \pm 20^\circ, 75^\circ \pm 15^\circ)$.  These galaxies are outliers.  \citet{aaro2} found the peculiar velocity field in the local supercluster is directed toward NGC 4486 (Messier 087) with $( l, b) \approx (284^\circ, 74^\circ)$ at a speed of 331$\pm$41 km~s$^{-1}$.  This has been called the ``Virgocentric infall''.  NGC 4486 is a peculiar, large, elliptical galaxy with strong X-ray emissions.  In addition, \citet{lilj} detected a quadrupolar tidal velocity field from spiral galaxy data in addition to the Virgocentric infall pointing toward (l,b) = (308$^\circ \, \pm 13^\circ $, 13$^\circ \, \pm 9^\circ $) at a speed of $\approx 200$ km~s$^{-1}$.  NGC 5128 (Centaurus A) at (l,b) = (310$^\circ$, 19$^\circ$) and at a distance of $ 3.84 \pm 0.35$ Mpc \cite{rejk} is a galaxy with properties \cite{isra} similar to NGC 4486.  \citet{lynd} found elliptical galaxies at distances in the 2000-7000 km~s$^{-1}$ range are streaming toward a ``Great Attractor'' centered on $( l, b) = (307^\circ \pm13^\circ, 9^\circ \pm8^\circ)$.  Centaurus B at (l,b) = (310$^\circ$, 2$^\circ$) is a galaxy with properties similar to NGC 4486.  In a more recent analysis, \citet{huds} suggested a bulk flow of 225 km~s$^{-1}$ toward $( l, b) \approx (300^\circ, 10^\circ)$.  However, the total mass in these directions appears to be insufficient to account for the peculiar velocity fields using Newtonian dynamics.

The changing universe model (CUM) proposed that Sources at the center of galaxies are continually erupting the Space and matter of our universe \cite{hodg4}.  Space is a constituent of our universe that ``flows'' from point Sources at the center of galaxies to Sinks.  The ``flow'' is like the nonviscous flow of heat from point sources to point sinks in a solid.  Space and the distribution of matter in Space are neither isotropic nor homogeneous.  The Space potential $\rho$ is like temperature in a solid.  The gradient $\bm \nabla \rho$ of the Space potential influences matter by exerting a force on the surface area of matter.  Photons are columns of basic particles called hods.  The hod is a two-dimensional surface with zero thickness.  As photons move through Space, hods may be stripped (redshift) from or added (blueshift) to the photon hod column.  

The $z - D$ relationship might show the reality of the physical nature of Space, of Space expansion, of the homogeneity of the universe, and of light.

This Paper examines the $z - D$ relationship of 32 spiral galaxies within 25 Mpc using the CUM of Space and of photons.  The developed equation results in a higher correlation coefficient than the Hubble law.  Also, a physical basis for the observed Virgocentric infall, for the Great Attractor phenomena, and for the Doppler shift of particle photons is given.

The object of this article is to apply the CUM beyond galaxy scale distances by developing an equation for the calculation of $z$ and by testing the equation using observations of galaxies that have $D$ calculated using Cepheid variable stars.  In section~\ref{sec:model}, the CUM $z$ calculation equation is developed.  The equation is applied to galaxies, and compared with galaxy observations in Section~\ref{sec:results}.  Some implications of the equation are discussed in~\ref{sec:impl}.  The results are discussed in Section~\ref{sec:disc}.  Section~\ref{sec:conc} lists the conclusions.

\section{\label{sec:model}MODEL}

The CUM suggests the energy of a photon is proportional to the number $N$ of hods in the particle column \cite{hodg4} of the photon.  The calculated redshift $z_\mathrm{c}$ is
\begin{equation}
z_\mathrm{c} \equiv \frac{\triangle \lambda}{\lambda} = \frac{N_\mathrm{e}}{N_\mathrm{o}}-1
\label{eq:3},
\end{equation}
where $\lambda$ is the wavelength, $N_\mathrm{e}$ is the emitted $N$, and $N_\mathrm{o}$ is the observed $N$.

From the energy continuity equation \cite{hodg4} of a volume enclosing the path of the photon in its travel from emitter to observer, $N$ is changed by the amount of Space through which the photon travels and the gravity of galaxies.  The loss of hods per unit volume is posited to be proportional the number of hods entering the volume $V$ of Space (Principle of Repetition),
\begin{equation}
\frac{\mathrm{d} N}{\mathrm{d} V} = -K_\mathrm{v} N
\label{eq:4},
\end{equation}
where $K_\mathrm{v}$ is the proportionality constant.

To explain phenomena of galaxies, the CUM also suggests hods combine to form photons only near a Source.  Therefore, there is a minimum number $N_\mathrm{min}$ of hods that a photon may have outside the immediate volume of the Source.  Combining Eqs.~(\ref{eq:3}) and (\ref{eq:4}) yields,
\begin{equation}
z_\mathrm{c} +1= \frac{N_\mathrm{e}}{N_\mathrm{min} + N_\mathrm{e} \, \exp({-K_\mathrm{v} \, \int_0^D \! \mathrm{d}V})}
\label{eq:6}.
\end{equation}

The $V$ is
\begin{equation}
V= C_\mathrm{s} \, D \, \bar{\rho}
\label{eq:7},
\end{equation}
where $C_\mathrm{s}$ is the cross section of $V$ and $\bar{\rho}$ is the average $\rho$ in $V$.

For the $D$ and change of $N$ considered herein, $C_\mathrm{s}$ is considered a constant.  For greater distances, where the total change of $N$ is relatively larger than considered herein, the $C_\mathrm{s}$ is a function of $N$ and $\rho$ at the position of the photon \cite{hodg4}.  The 
\begin{equation}
\bar{\rho} = \frac{1}{D} \, \int_0^D \rho\, \mathrm{d}x
\label{eq:8},
\end{equation}
where d$x$ is the incremental ho distance $x$ traveled by the photon.

The conservation of matter, which includes photons as matter particles, implies for each galaxy
\begin{equation}
\epsilon_{\mathrm{m}} + K_\mathrm{lm} L_{\mathrm{I}} + M_{\mathrm{I}} = K_\mathrm{lm} L_{\mathrm{O}} +M_{\mathrm{O}} + M_{\mathrm{g}}
\label{eq:9},
\end{equation}
where $\epsilon_{\mathrm{m}}$ ($M_\odot$~s$^{-1}$), $M_{\mathrm{I}}$ ($M_\odot$~s$^{-1}$), $M_{\mathrm{O}}$ ($M_\odot$~s$^{-1}$), $M_{\mathrm{g}}$ ($M_\odot$~s$^{-1}$), $L_{\mathrm{I}}$ (erg~s$^{-1}$), $L_{\mathrm{O}}$ (erg~s$^{-1}$), and $K_\mathrm{lm}$ ($M_\odot$/erg) are the amount of mass emitted per second by the Source of the galaxy, the matter per second (exclusive of photons) into the galaxy from other galaxies, the matter per second (exclusive of photons) emitted by the galaxy, the matter per second increase in the galaxy, the luminosity due to photons into the galaxy, the luminosity due to photons emitted from the galaxy and the proportionality constant to convert erg to $M_\odot$, respectively.  Matter is ``in'' a galaxy when it is in orbit around the Source, where the dynamics of the matter is principally determined by the galaxy parameters.

The feedback mechanism at the center of galaxies causes the mass of a mature and stable galaxy to remain constant \cite{hodg4}.  Therefore, $ M_{\mathrm{g}} = 0$ and the mass of a mature galaxy is proportional to the $\epsilon_{\mathrm{m} }$.  Therefore, the net effect of the mass and Space forces of the galaxy is proportional to $\epsilon_{\mathrm{m} }$.  The $\rho$ at a point $i$ in Space ($\rho_i$) is the sum of the $\rho$ effect of all galaxies \cite{hodg4},
\begin{subequations}
\label{eq:13}
\begin{eqnarray}
\rho_i = & K_\mathrm{\rho} \, \sum_{k =1}^\mathrm{all \, galaxies} \, \frac{\epsilon_{\mathrm{m} k} }{r_{ik}} \, & \mathrm{if}  \sum_{k =1} ^\mathrm{all \, galaxies} \, \frac{\epsilon_{\mathrm{m} k} }{r_{ik}}>0, \label{eq:13a}\\*
\rho_i = & 0 \, & \mathrm{if} \sum_{k =1}^\mathrm{all \, galaxies} \, \frac{\epsilon_{\mathrm{m} k} }{r_{ik}} \leq 0, \label{eq:13b}
\end{eqnarray}
\end{subequations}
where the roman, italic subscript denotes an index of the parameter to which it is a subscript, $ K_\mathrm{\rho}$ is a proportionality constant, and $r_{ik}$ is the ho distance from the $k^{\mathrm{th}}$ galaxy to the $i^{\mathrm{th}}$ point.  The $\epsilon_{\mathrm{m} } >0$ for Sources and $ \epsilon_{\mathrm{m} }<0$ for Sinks. 

A dilemma is created by considering the $\epsilon_{\mathrm{m} } < 0$ for a Sink.  The calculation of Eq.~(\ref{eq:13a}) will yield $\rho_i <0$ for regions near Sinks.  The CUM proposes the $\rho$ must be positive or, in a void, zero.  The $\epsilon_{\mathrm{m} }$ of a Sink is the amount of mass ejected from our universe.  Therefore, to overcome this calculation dilemma, in regions where Eq.~(\ref{eq:13a}) yields a $\rho_i <0$, the values of $\rho_i =0$, $\bm \nabla \rho_i =0$, and $\nabla^2 \rho_i =0$ will be used.

The CUM proposes that in a galaxy with no neighbor galaxies (an intrinsic galaxy), the mass emitted by the Source remains around the Source.  Since Space is causally correlated and coherent \cite{hodg4}, the presence of other galaxies causes a $\bm{\nabla} \rho$ that can cause matter to be removed from orbiting a galaxy.  Other galaxies or Sinks might capture the matter removed from a galaxy's orbit.  The net force on a particle from a Source type galaxy is the \emph{difference} of the $\bm{\nabla} \rho$ force and gravity.  Thus, the effective mass of Source type galaxy appears less than Newtonian expectation.  Conversely, the net force on a particle from a Sink type galaxy is the \emph{sum} of the $\bm{\nabla} \rho$ force and gravity.  Thus, the effective mass of a Sink type galaxy appears more than Newtonian expectation.  For a photon, the cross section perpendicular to the column of hods is zero.  Therefore, the $\bm{\nabla} \rho$ force perpendicular to the column is zero.  Therefore, the photon experiences the full gravitational force whereas baryonic particles experience the gravitational force modified by the $\bm{\nabla} \rho$ force.

This Paper posits the net mass removed from the galaxy is destined to be captured by a Sink, only, and is proportional (Principle of Feedback) to the $\bm{\nabla} \rho$ at the center of the galaxy due to all other galaxies ($\bm{\nabla} \rho_{\mathrm{c} } $),
\begin{equation} 
M_{\mathrm{O}} - M_{\mathrm{I}} \approx K_\mathrm{s \epsilon} \vert \bm{\nabla} \rho_{\mathrm{c} } \vert 
\label{eq:10},
\end{equation}
where $K_\mathrm{s \epsilon}$ is a proportionally constant and $\vert \bm{\nabla} \rho_{\mathrm{c} } \vert$ is the value of $\bm{\nabla} \rho_{\mathrm{c} } $. 

The $L_{\mathrm{O}}$ of a galaxy is from photons emitted by the mass and the Source of the galaxy.  The $L_{\mathrm{O}}$ of a galaxy is assumed to be isotropic.  Other factors such as the K-correction were ignored.  The $L_{\mathrm{O}}$ is posited to be proportional to the flux in the $\beta$-band at a standard distance from the galaxy.  Therefore, to a first approximation, 
\begin{equation}
\epsilon_{\mathrm{m} } = K_\mathrm{lm} L_{\mathrm{O} } + K_\mathrm{s \epsilon} \vert \bm{\nabla} \rho_{\mathrm{c} } \vert 
\label{eq:11},
\end{equation}
where
\begin{equation}
K_\mathrm{lm} L_{\mathrm{O}}=K_\mathrm{lf} 10^{-0.4 \, M_{\beta }}
\label{eq:12};
\end{equation}
\begin{equation}
 M_{\beta } \equiv m_{\beta } -E_{\mathrm{xt} } +25 - 5 \, \log(D)
\label{eq:12a};
\end{equation}
$ K_\mathrm{lf}$ is a proportionality constant; and $M_{\beta }$ (mag.), $m_{\beta }$ (mag.), and $E_{\mathrm{xt} }$ (mag.) are the absolute magnitude, apparent magnitude, and extinction, respectively, of the galaxy. 

Since $V$ is a function of $x$ and time $t$,
\begin{equation}
\mathrm{d}V(x,t)=\frac{\partial V}{\partial x}\mathrm{d} x + \frac{\partial V}{\partial t}\mathrm{d} t
\label{eq:14}.
\end{equation}

Expressing Eq.~(\ref{eq:14}) as a function of $\rho$ and $D$ and substituting into Eq.~(\ref{eq:6}) yields,
\begin{eqnarray}
\frac{1}{z_\mathrm{c}+1}&=& K_\mathrm{min} + \exp( K_\mathrm{dp} D P + K_\mathrm{d} D +K_\mathrm{p} P \nonumber \\*
& &+ K_\mathrm{f} F + K_\nabla \int_0^D \! \nabla ^2 \rho \, \mathrm{d} x + K_\mathrm{vp} P v_\mathrm{e}
\label{eq:15},
\end{eqnarray}
where: (1) Relatively small terms such as terms involving the relative $\epsilon$ of the emitter and observer were ignored.
(2)  The $ K_\mathrm{min} =N_\mathrm{min}/N_\mathrm{e}$ is a constant for a given $N_\mathrm{e}$.
(3)  The $K_\mathrm{dp}$, $K_\mathrm{d}$, $K_\mathrm{p}$, $K_\mathrm{f}$, $K_\nabla$, and $K_\mathrm{vp}$ are constants.
(4)  The $P= \int_0^D \! \rho_\mathrm{wo} \mathrm{d} x$, where $\rho_\mathrm{wo}$ is the $\rho_i$ calculated from Eq.~(\ref{eq:13}) with the emitter and observer galaxies omitted.
(5)  The $F = \int_0^D \! [(\partial \rho / \partial x)-K_\mathrm{co}] \mathrm{d} x$, where $K_\mathrm{co}$ is the minimum $(\partial \rho / \partial x)$ that removes hods from the photon and the Principle of Feedback has been used to derive the linear relationship.
(6)  The $\nabla ^2 \rho$ term derives from $\mathrm{d} \rho/\mathrm{d} t =\nabla ^2 \rho $ of the energy continuity equation.
(7)  The relative velocity of the emitting and observing galaxies causes a change in $V$, hence $N$, and has three causes.  One is the expansion of our universe due to expansion of the second dimension \cite{hodg4}.  By the Principle of Negative Feedback, this component is linearly related to the ho distance $(K_\mathrm{dp} P + K_\mathrm{d} ) D$, where $K_\mathrm{dp}$ and $K_\mathrm{d}$ are constants.  The second cause is due to the possible peculiar velocity of the Milky Way relative to the reference frame derived by summing over all Sources and Sinks similar to Mach's principle \cite{hodg4}.  Another cause derives from the inaccuracy of defining the reference frame because the Sources and Sinks directly on the other side of the Milky Way center from earth are unobservable from earth.  The component $v_\mathrm{e}$ deriving from the second and third causes is proportional to the cosine of the angular difference between the target galaxy and the direction of $v_\mathrm{e}$.  Thus,

\begin{eqnarray}
v_\mathrm{e} &=& \cos(90^\circ - L_\mathrm{at}) \, \cos(90^\circ -K_\mathrm{lat}) \, \nonumber \\*
& &+ \, \sin(90^\circ - L_\mathrm{at}) \, \sin(90^\circ -K_\mathrm{lat}) \, \nonumber \\
& & \times \, \cos(L_\mathrm{on}-K_\mathrm{lon})
\label{eq:16},
\end{eqnarray}
where $L_\mathrm{at}$ (degrees) and $L_\mathrm{on}$ (degrees) are the galactic latitude and longitude, respectively, of the emitting galaxy; and $K_\mathrm{lat}$ and $K_\mathrm{lon}$ are the galactic latitude and galactic longitude, respectively, of the direction of $v_\mathrm{e}$.

\section{\label{sec:results}Results}

The sample galaxies were selected from the NED database\footnote{The Ned database is available at http::/nedwww.ipac.caltech.edu.  The data were obtained from NED on 5 May 2004.}.  The selection criteria were that the heliocentric redshift $z_\mathrm{mh}$ be less than 0.03 and that the object be a galaxy.  The parameters obtained from the NED database included the galaxy name, $L_\mathrm{on}$, $L_\mathrm{at}$, the heliocentric redshift $z_\mathrm{mh}$, morphology, the $m_{\mathrm{\beta}}$ in the visible range as defined by NED, and the galactic extinction $E_{\mathrm{xt}}$.  The $z_{\mathrm{m}}$ was calculated from $z_\mathrm{mh}$.

The 21-cm line width $W_\mathrm{20}$ (km~s$^{-1}$) at 20 percent of the peak and the inclination $i_\mathrm{n}$ (degrees) between the line of sight and polar axis were obtained from the LEDA database\footnote{The LEDA database is available at http://leda.univ-lyon.fr.  The data were obtained from LEDA on 5 May 2004.} when such data existed.

In the CUM, peculiar velocity is caused by $P$ and $F$.  Therefore, the inward peculiar velocity field of the Virgocentric infall is because NGC 4486, NGC 5128, and Centaurus B were considered Sinks with mass surrounding them.  That additional Sinks may exist in the direction of (l,b) = (300$^\circ$, 10$^\circ$) has also been suggested \cite{huds}.

The constants to be discovered are the constants of Eqs.~(\ref{eq:11}) and (\ref{eq:15}), the luminosity of the three Sinks, the distance to NGC 5128 and Centaurus B, and the luminosity of the Milky Way.  Since the $D$ and $\epsilon$ of all galaxies consistent with the CUM must also be found, calculating the constants was done by making several simplifying assumptions and by iteration.  The simplifying assumptions were: (1) The distance $D_\mathrm{a}$ to the 32 trial galaxies calculated by \citet{free} and \citet{macr} using Cepheid variable stars are proportional to the CUM ho distance\cite{hodg4}.  (2) The $r_{ik}$ is limited to 30 Mpc except for the Sinks to reduce the selection bias caused by limiting the $z_\mathrm{m}$ of the galaxies selected.  (3) Galaxies with an unlisted morphology in the NED database were ignored.  An alternate method may be to assign a low value of $m_{\mathrm{\beta}}$ to such galaxies.  This option was rejected because a large number of such galaxies were from the 2dFGRS and 2MASS.  These surveys include only limited areas of the sky.  Therefore, including the 2dFGRS and 2MASS galaxies with unlisted morphology in the calculation would introduce a selection bias into the sample.  (4) Galaxies with an unlisted $m_{\beta }$ were assigned $ m_{\beta } = -11$ mag.  (5)  Objects with $E_{\mathrm{xt}} = 99$ were assigned an $E_{\mathrm{xt}k}=0$ mag.  (6)  All the sample galaxies were considered mature and stable ($M_{\mathrm{g}}=0$).  The result was 29,984 sample galaxies.

The iteration procedure was: (1) Estimate the $D$ to the sample galaxies (see Appendix~\ref{sec:initial}).  (2) Calculate the $\epsilon_{\mathrm{m} }$ of the sample galaxies using $D$ and Eq.~(\ref{eq:11}).  (3) Using the $D$ and $\epsilon_{\mathrm{m} }$ of the galaxies, calculate the $F$, $P$, and $\nabla^2 \rho$ of the trial galaxies.  (4) For the trial galaxies, adjust the constants, calculate the redshift $z_\mathrm{c}$ of Eq.~(\ref{eq:15}), and calculate
\begin{equation}
z_\mathrm{c} = K_\mathrm{scm} z_\mathrm{m} + K_\mathrm{icm}
\label{eq:25},
\end{equation}
where $K_\mathrm{scm}$ is the least squares slope and $K_\mathrm{icm}$ is the least squares intercept of the presumed linear relationship.  Adjust the constants to maximize the correlation coefficient of Eq.~(\ref{eq:25}) with $K_\mathrm{scm} \approx 1$ and with $ K_\mathrm{icm} \approx 0$.  (5) Calculate the redshift $z_\mathrm{cs}$ of the sample galaxies with $D < 45$ Mpc using $z_\mathrm{cs} = z_\mathrm{c}$ of Eq.~(\ref{eq:15}), the constants found in step (4), and the current value of $D$ for each galaxy.  (6) If $z_\mathrm{cs} < (1- 0.6 K_\mathrm{cm}) z_\mathrm{m} $, where $0 < K_\mathrm{cm} < 1$ is an assigned constant, then the new estimated distance $D_\mathrm{new} = (1+ K_\mathrm{cm}) D$.  If $z_\mathrm{cs} > (1+ 0.6 K_\mathrm{cm}) z_\mathrm{m} $, then $D_\mathrm{new} = (1- K_\mathrm{cm}) D$. (7) Set $D= D_\mathrm{new}$ for the Category C through Category E sample galaxies that have a $D < 45$ Mpc.  (8) repeat steps (2) through (8).  For the first, second, and third iteration, $ K_\mathrm{cm} = $ 0.15, 0.10, and 0.10, respectively.

\begin{figure*}
\includegraphics[width=\textwidth]{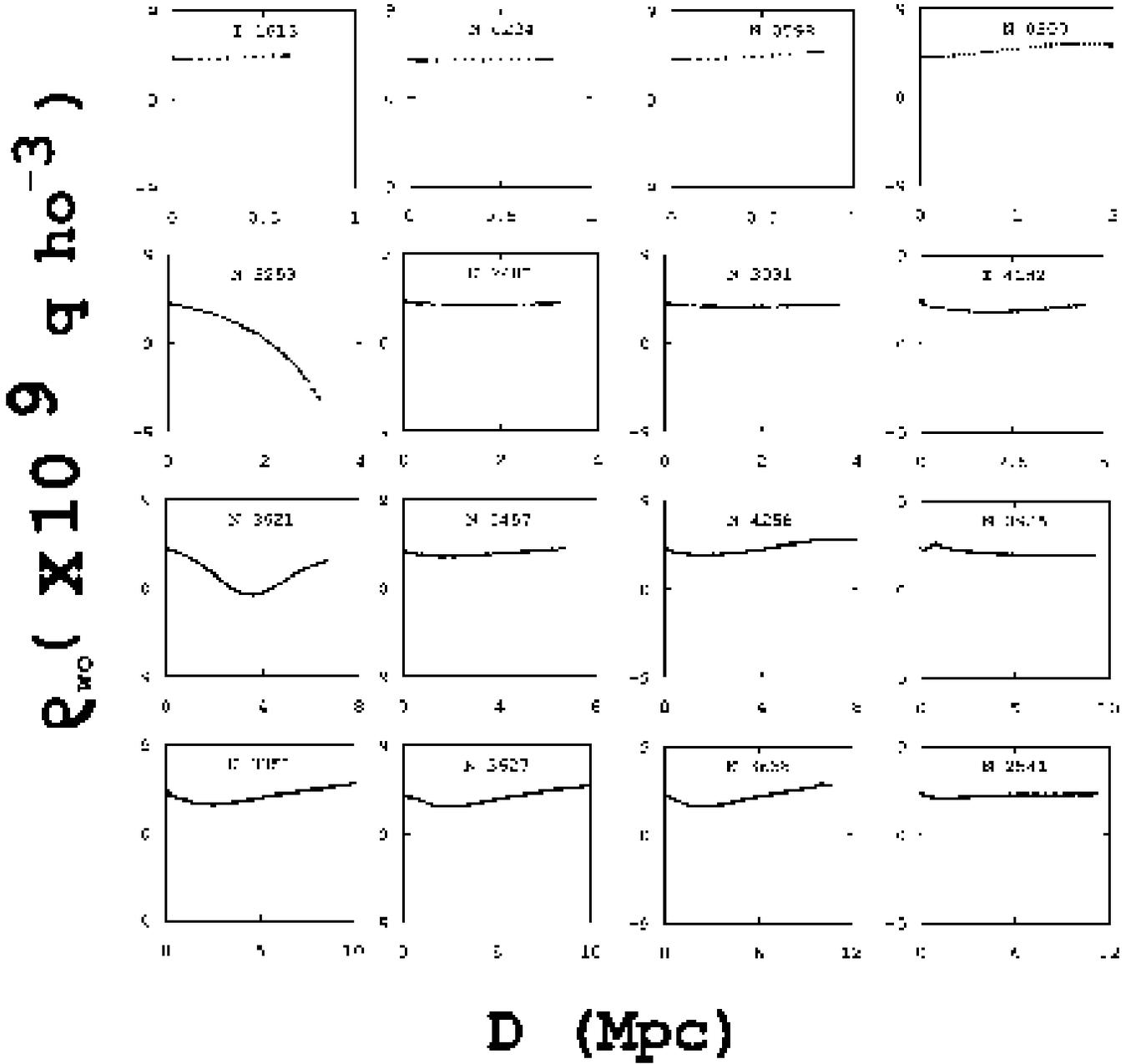}
\caption{\label{fig:6}Plot of $\rho_\mathrm{wo} \, (\times 10^{-3} $ q~ho$^{-3}$) versus $D_\mathrm{a}$ (Mpc) of the Category A galaxies.}
\end{figure*}

Figure~\ref{fig:6} shows plots of $\rho_\mathrm{wo} \, (\times 10^{-3} $ q~ho$^{-3}$) versus $D_\mathrm{a}$ (Mpc) for the Category A galaxies.  

\addtocounter{figure}{-1}
\begin{figure*}
\includegraphics[width=\textwidth]{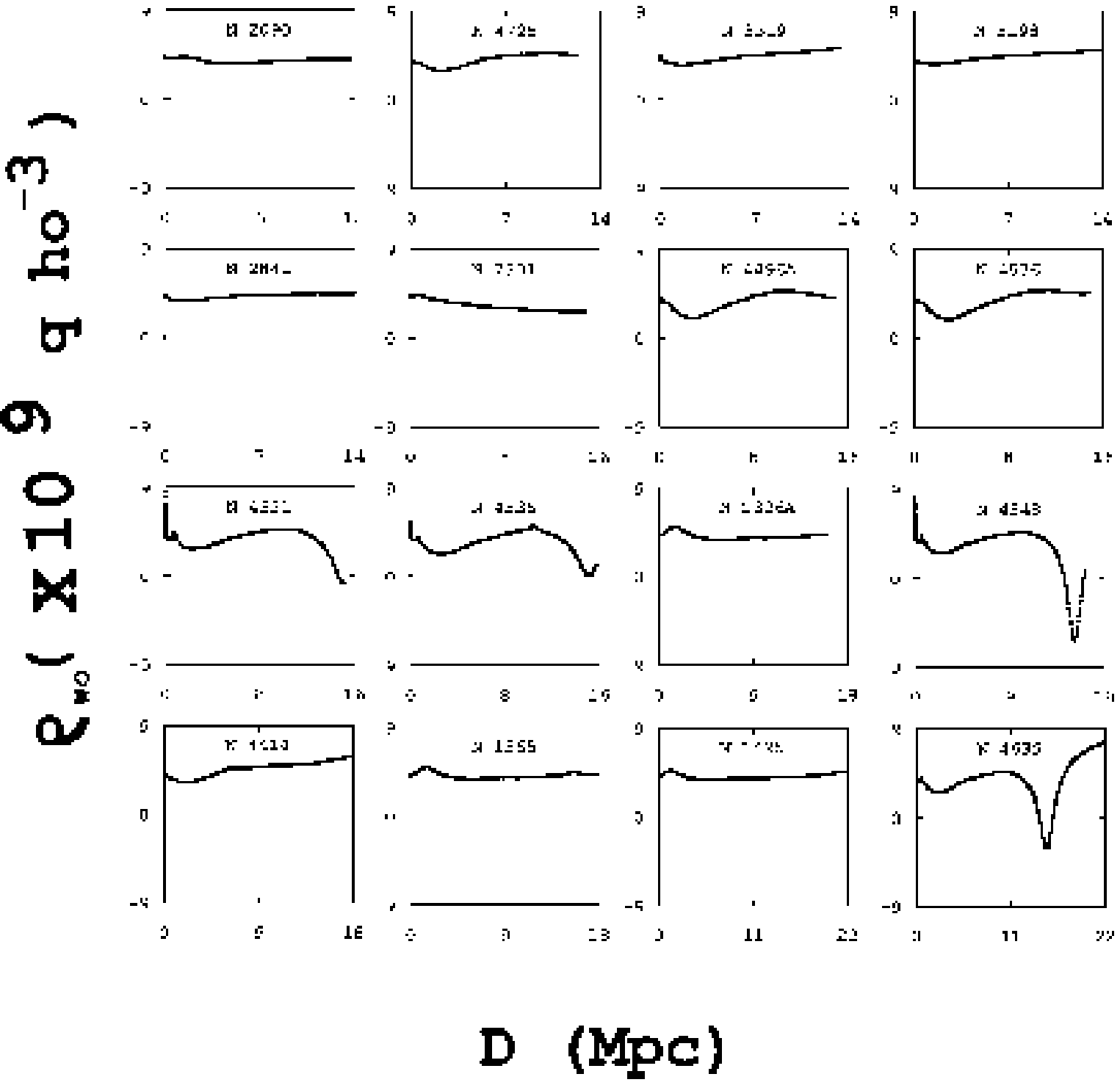}
\caption{\label{fig:7}(Cont.) }
\end{figure*}

\begin{figure}
\includegraphics[width=0.5\textwidth]{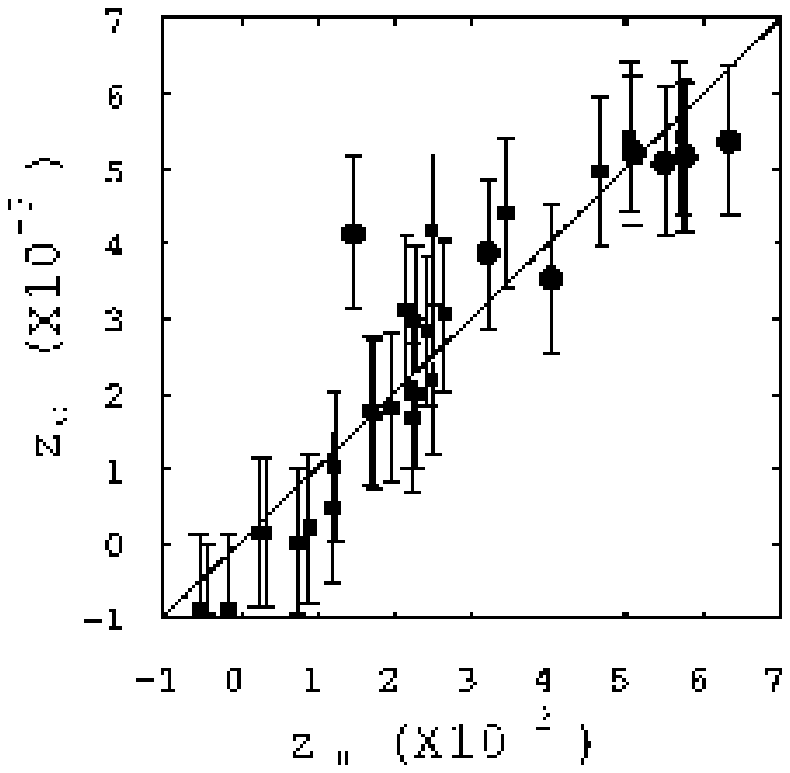}
\caption{\label{fig:4} Plot of the measured redshift $z_\mathrm{m}$ versus the calculated redshift $z_\mathrm{c}$ using Eq.~(\ref{eq:15}) for 32 Category A sample galaxies \cite{free,macr}.  The straight line indicates $z_\mathrm{c} = z_\mathrm{m}$.  The circles indicate the data points for galaxies with (l,b) = (290$^\circ \pm 20^\circ$,75$^\circ \pm 15^\circ$).}
\end{figure}

Figure~\ref{fig:4} shows a plot of $z_\mathrm{c}$ versus $z_{\mathrm{m}}$ for the trial galaxies after the third iteration.  Tables~\ref{tab:1}, \ref{tab:2}, and \ref{tab:4} lists the data for the trial galaxies.  After the third iteration, $K_\mathrm{scm} = 1.02 \pm 0.08$ and $K_\mathrm{icm}= (0 \pm 6) \times 10^{-5} $ at 1$\sigma$.  The correlation coefficient is 0.93.

\begingroup
\squeezetable
\begin{table*}
\caption{\label{tab:1} Trial galaxy data in order of increasing $D_\mathrm{a}$ after the third iteration. }
\begin{ruledtabular}
\begin{tabular}{llrrrrrrrrr}
{Galaxy}
&{Morphology}
&$L_\mathrm{on}$\footnotemark[1]
&{$L_\mathrm{at}$\footnotemark[1]}
&{$D_\mathrm{a}$} 
&{$z_\mathrm{m}$}
&{$z_\mathrm{h}$}
&{$z_\mathrm{c}$ }
&{$P$}
&{$F$}
&{$\int_0^D \! \nabla ^2 \rho \, \mathrm{d} x $} \\
& &&&& $\times 10^{-3}$&$\times 10^{-3}$ &$\times 10^{-3}$ & $\times 10^{10}$& $\times 10^{12}$ & $\times 10^{8}$ \\
& &deg.&deg.&Mpc&& & & q~Mpc$^{-2}$&q~Mpc$^{-3}$& q~Mpc$^{-2}$\\
\hline
IC 1613&IB(s)m&130&-61&0.65&-0.518&0.152&-0.872&0.27&0.07&0.30\\
NGC 0224&SA(s)b LINER&121&-22&0.79&-0.408&0.184&-1.008&0.30&0.06&-1.54\\
NGC 0598&SA(s)cd HII&134&-31&0.84&-0.148&0.196&-0.866&0.36&0.08&0.30\\
NGC 0300&SA(s)d&299&-79&2.00&0.336&0.467&0.171&0.97&0.21&0.82\\
NGC 5253&Im pec;HII Sbrst&315&30&3.15&0.903&0.735&0.215&0.52&0.22&-0.54\\
NGC 2403&SAB(s)cd HII&151&29&3.22&0.758&0.751&0.013&1.24&0.33&-0.08\\
NGC 3031&SA(s)ab;LINER Sy1.8&142&41&3.63&0.243&0.847&0.152&1.37&0.36&-1.82\\
IC 4182&SA(s)m  &108&79&4.49&1.231&1.048&1.042&1.58&0.47&0.60\\
NGC 3621&SA(s)d  &281&26&6.64&1.759&1.549&1.746&0.96&0.51&0.42\\
NGC 5457&SAB(rs)cd  &102&60&6.70&1.202&1.563&0.482&2.34&0.47&2.22\\
NGC 4258&SAB(s)bc;LINER Sy1.9 &138&69&7.98&1.694&1.862&1.770&3.32&0.81&1.86\\
NGC 0925&SAB(s)d  HII     &145&-25&9.16&2.216&2.137&2.028&3.43&0.95&-0.06\\
NGC 3351&SB(r)b;HII  Sbrst   &234&56&10.00&2.258&2.333&2.987&3.75&1.03&0.00\\
NGC 3627&SAB(s)b;LINER  Sy2     &242&64&10.05&2.145&2.345&3.123&3.73&1.03&0.35\\
NGC 3368&SAB(rs)ab;Sy  LINER   &234&57&10.52&2.659&2.455&3.050&4.01&1.08&-0.19\\
NGC 2541&SA(s)cd  LINER   &170&33&11.22&1.963&2.618&1.801&4.45&1.16&0.22\\
NGC 2090&SA (rs)b  &239&-27&11.75&2.490&2.742&4.188&4.50&1.22&-0.27\\
NGC 4725\footnotemark[3]&SAB(r)ab pec  Sy2     &295&88&12.36&4.026&2.884&3.528&4.95&1.26&7.18\\
NGC 3319&SB(rs)cd  HII     &176&59&13.30&2.497&3.103&2.196&5.75&1.38&0.32\\
NGC 3198&SB(rs)c  &171&55&13.80&2.281&3.220&2.016&5.93&1.43&0.13\\
NGC 2841&SA(r)b ;LINER  Sy1     &167&44&14.07&2.249&3.283&1.688&5.80&1.42&-1.93\\
NGC 7331&SA(s)b  LINER   &94&-21&14.72&3.434&3.435&4.394&4.75&1.51&2.25\\
NGC 4496A\footnotemark[2]&SB(rs)m  &291&66&14.86&5.505&3.467&5.109&5.46&1.54&-1.03\\
NGC 4536\footnotemark[2]&SAB(rs)bc      HII  &293&65&14.93&5.752&3.484&5.177&5.50&1.55&-0.25\\
NGC 4321\footnotemark[2]&SAB(s)bc;LINER HII  &271&77&15.21&5.087&3.549&5.227&5.35&1.53&53~440.13\\
NGC 4535\footnotemark[2]&SAB(s)c        HII  &290&71&15.78&6.325&3.682&5.382&5.21&1.63&-8.28\\
NGC 1326A&SB(s)m  &239&-56&16.14&5.713&3.766&5.401&6.53&1.68&0.28\\
NGC 4548\footnotemark[2]&SBb(rs);LINER  Sy      &286&77&16.22&1.476&3.785&4.144&5.17&1.47&192.50\\
NGC 4414&SA(rs)c?  LINER   &175&83&17.70&2.416&4.130&2.835&8.04&1.82&4.14\\
NGC 1365&(R')SBb(s)b    Sy1.8  &238&-55&17.95&5.049&4.188&5.424&7.29&1.85&-0.67\\
NGC 1425&SA(rs)b  &228&-53&21.88&4.666&5.105&4.951&8.89&2.26&-0.36\\
NGC 4639\footnotemark[2]&SAB(rs)bc  Sy1.8   &294&76&21.98&3.223&5.129&3.853&8.54&2.11&82.26\\
\end{tabular}
\end{ruledtabular}

\footnotetext[1]{Rounded to the nearest degree. }
\footnotetext[2]{The data points for galaxies with (l,b) = (290$^\circ \pm 20^\circ$,75$^\circ \pm 15^\circ$) (see Figs.~\ref{fig:1} and \ref{fig:4}).}
\end{table*}
\endgroup

\begingroup
\squeezetable
\begin{table*}
\caption{\label{tab:2} The components of Eq.~(\ref{eq:15}) for each trial galaxy after the third iteration. }
\begin{ruledtabular}
\begin{tabular}{lrrrrrr}
{Galaxy}
&{$ K_\mathrm{dp} D P $}
&{$ K_\mathrm{p} P $}
&{$ K_\mathrm{f} F $}
&{$ K_\nabla \int_0^D \! \nabla ^2 \rho \, \mathrm{d} x$} 
&{$ K_\mathrm{vp} P v_\mathrm{e}$}
&{$X$}\\ 
&$\times 10^{-3}$&$\times 10^{-3}$&$\times 10^{-3}$&$\times 10^{-9}$&$\times 10^{-3}$&$\times 10^{-3}$\\
\hline
IC 1613&0.007&-0.03&-0.40&-6.31&0.028&-0.396\\
NGC 0224&0.010&-0.04&-0.33&32.44&0.102&-0.261\\
NGC 0598&0.013&-0.05&-0.49&5.59&0.120&-0.403\\
NGC 0300&0.081&-0.12&-1.20&-17.15&-0.202&-1.441\\
NGC 5253&0.069&-0.07&-1.31&9.85&-0.179&-1.485\\
NGC 2403&0.168&-0.16&-1.93&1.76&0.636&-1.283\\
NGC 3031&0.209&-0.18&-2.09&38.23&0.637&-1.422\\
IC 4182&0.298&-0.20&-2.72&-12.56&0.313&-2.313\\
NGC 3621&0.267&-0.12&-2.97&-8.84&-0.190&-3.016\\
NGC 5457&0.657&-0.30&-2.73&-46.69&0.624&-1.753\\
NGC 4258&1.111&-0.43&-4.73&-39.05&1.006&-3.041\\
NGC 0925&1.315&-0.44&-5.53&1.17&1.352&-3.299\\
NGC 3351&1.570&-0.48&-6.01&-0.07&0.668&-4.256\\
NGC 3627&1.570&-0.48&-6.02&-7.33&0.534&-4.392\\
NGC 3368&1.766&-0.51&-6.28&3.92&0.704&-4.320\\
NGC 2541&2.090&-0.57&-6.78&-4.70&2.185&-3.071\\
NGC 2090&2.215&-0.58&-7.08&5.59&-0.016&-5.455\\
NGC 4725&2.564&-0.63&-7.35&-150.80&0.625&-4.796\\
NGC 3319&3.202&-0.74&-8.03&-6.81&2.096&-3.466\\
NGC 3198&3.426&-0.76&-8.31&-2.77&2.358&-3.287\\
NGC 2841&3.419&-0.74&-8.29&40.48&2.653&-2.959\\
NGC 7331&2.930&-0.61&-8.77&-47.24&0.792&-5.660\\
NGC 4496A\footnotemark[1]&3.401&-0.70&-8.99&21.55&-0.087&-6.373\\
NGC 4536\footnotemark[1]&3.440&-0.70&-9.00&5.18&-0.182&-6.441\\
NGC 4321\footnotemark[1]&3.408&-0.68&-8.93&-1~122~242.83&0.461&-6.864\\
NGC 4535\footnotemark[1]&3.445&-0.67&-9.49&173.88&0.071&-6.645\\
NGC 1326A&4.416&-0.84&-9.77&-5.87&-0.472&-6.663\\
NGC 4548\footnotemark[1]&3.510&-0.66&-8.57&-4~042.45&0.313&-5.412\\
NGC 4414&5.962&-1.03&-10.60&-86.90&1.561&-4.104\\
NGC 1365&5.480&-0.93&-10.76&14.03&-0.475&-6.686\\
NGC 1425&8.150&-1.14&-13.18&7.60&-0.047&-6.215\\
NGC 4639\footnotemark[1]&7.869&-1.09&-12.25&-1~727.43&0.359&-5.121\\
\end{tabular}
\end{ruledtabular}

\footnotetext[1]{The data points for galaxies with (l,b) = (290$^\circ \pm 20^\circ$,75$^\circ \pm 15^\circ$) (see Figs.~\ref{fig:1} and \ref{fig:4}).}

\end{table*}
\endgroup

\begin{table*}
\caption{\label{tab:4} The values of the constants of Eq.~(\ref{eq:15}) after the third iteration. }
\begin{ruledtabular}
\begin{tabular}{lrl}
{Parameter}
&{value}
&units\\ 
\hline
Milky Way Luminosity&$1.58 \times 10^7$&$\mathrm{erg} \, \mathrm{s}^{-1}$\\
NGC 4486 Luminosity&$-9.01 \times 10^9$&$\mathrm{erg} \, \mathrm{s}^{-1}$\\
NGC 4486 Distance&15.2\phantom{0}&Mpc\\
NGC 5128 Luminosity&$-1.21 \times 10^{10}$&$\mathrm{erg} \, \mathrm{s}^{-1}$\\
NGC 5128 Distance\footnotemark[1]&3.85&Mpc\\
PKS 1343-60 Luminosity&$-6.21 \times 10^9$&$\mathrm{erg} \, \mathrm{s}^{-1}$\\
PKS 1343-60 Distance&22.0\phantom{0}&Mpc\\
$K_\mathrm{min}$&$1.269 \times 10^{-3}$&\\
$K_\mathrm{dp}$&$4.190 \times 10^{-15}$&q$^{-1}$~ho$^{3}$\\
$K_\mathrm{d}$\footnotemark[2]&$0.00$&Mpc$^{-1}$\\
$K_\mathrm{p}$&$-1.280 \times 10^{-14}$&q$^{-1}$~Mpc$^{-1}$~ho$^{3}$\\
$K_\mathrm{f}$&$-5.820 \times 10^{-15}$&q$^{-1}$~ho$^{3}$\\
$K_\mathrm{co}$&$-1.040 \times 10^{11}$&q~ho$^{-3}$~Mpc$^{-1}$\\
$K_\nabla$&$4.190 \times 10^{-15}$&q$^{-2}$~ho$^{6}$~Mpc\\
$K_\mathrm{vp}$&$5.300 \times 10^{-14}$&q$^{-1}$~Mpc$^{-1}$~ho$^{3}$~deg$^{-1}$\\
$L_\mathrm{at}$&$15$&$^\circ$\\
$L_\mathrm{on}$&$157$&$^\circ$\\
\end{tabular}
\end{ruledtabular}
\footnotetext[1]{From reference \cite{rejk}.}
\footnotetext[2]{Since $K_\mathrm{d}=0$, it is ignored in the final equation.}

\end{table*}

The $K_\mathrm{d} \approx 0$ and, therefore, was ignored.  Therefore, Eq.~(\ref{eq:15}) becomes 
\begin{equation}
\frac{1}{z_\mathrm{c}+1}= K_\mathrm{min} + \mathrm{e}^X
\label{eq:26},
\end{equation}
where
\begin{equation}
X= K_\mathrm{dp} D P + K_\mathrm{p} P + K_\mathrm{f} F + K_\nabla \int_0^D \! \nabla ^2 \rho \, \mathrm{d} x + K_\mathrm{vp} P v_\mathrm{e}
\label{eq:27}.
\end{equation}

\begin{figure}
\includegraphics[width=0.5\textwidth]{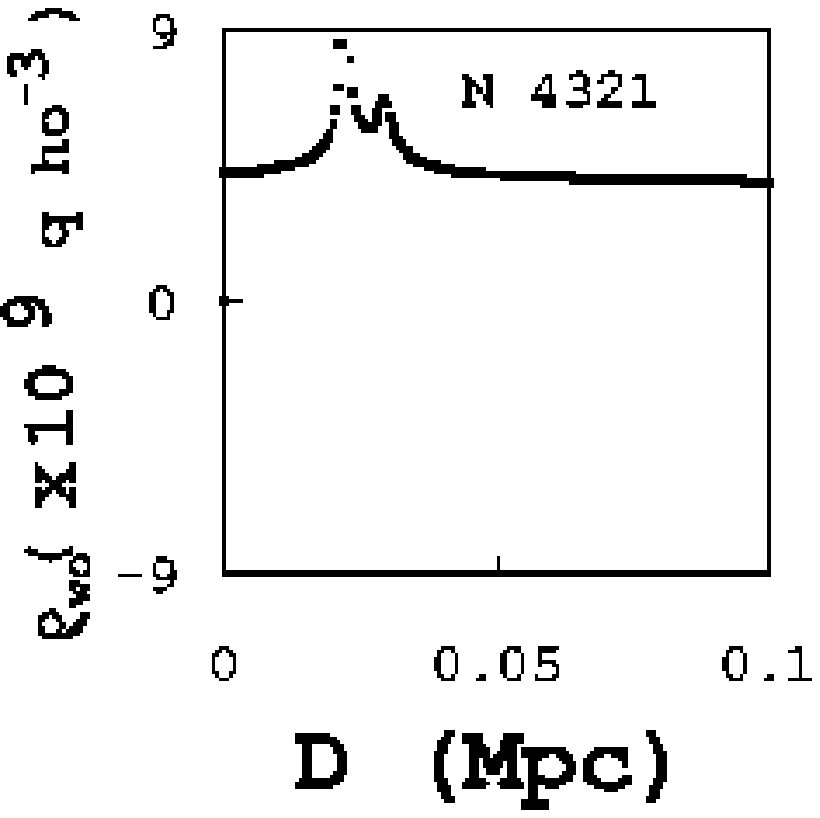}
\caption{\label{fig:8}Plot of $\rho_\mathrm{wo} \, (\times 10^{-3} $ q~ho$^{-3}$) versus $D_\mathrm{a}$ (Mpc) of NGC 4321 for $D_\mathrm{a} < 0.5$ Mpc.}
\end{figure}

The values of the $ K_\nabla \int_0^D \! \nabla ^2 \rho \, \mathrm{d} x$ term are small except for NGC 4321.  As seen in Fig.~\ref{fig:8}, The photon path passes very near galaxies at 0.02 Mpc and 0.03 Mpc.  This causes the $ K_\nabla \int_0^D \! \nabla ^2 \rho \, \mathrm{d} x$ term to be greater than other Category A galaxies.

\section{\label{sec:impl}Implications}

The $K_d =0$.  The Doppler shift is independent of a distance, only, term.  Rather, the Doppler shift is dependant on distance because a larger distance implies larger $P$ and $F$ terms.  The Doppler shift observed in particle photons is partially caused by the rate of increasing amount of Space between emitter and observer.  For values of $D$ of the order of size of the solar system and for volumes in a galaxy, $P \approx 0$, $F \approx 0$, and the $\nabla^2 \rho = \mathrm{d} \rho / \mathrm{d} t$ term predominates and is interpreted to be the relative velocity $v$ between emitter and observer.  Thus, $z \propto v$.  The Newtonian redshift is recovered.

\begin{figure}
\includegraphics[width=0.5\textwidth]{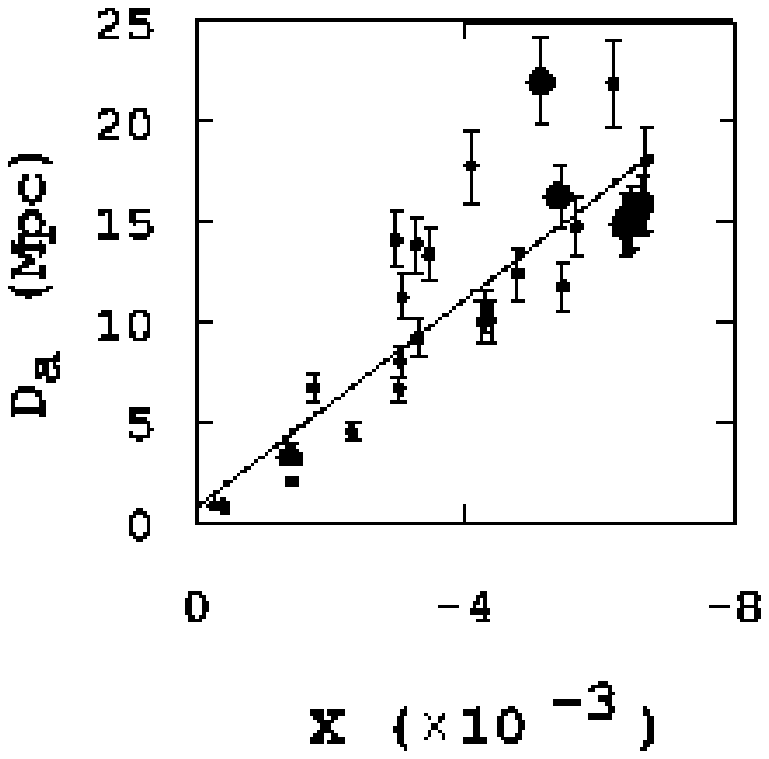}
\caption{\label{fig:5}Plot of $D_\mathrm{a}$ (Mpc) versus $X$.  The larger circles are the data points for the galaxies with (l,b) = (290$^\circ \pm 20^\circ$,75$^\circ \pm 15^\circ$) (see Figs.~\ref{fig:1} and \ref{fig:4}).}
\end{figure}

The analysis herein calculated $D$ of galaxies within approximately 25 Mpc from earth.  Assuming the volumes of space with radii in increments of 50 Mpc are approximately as the local volume, the $X$ term will become predominate and $K_\mathrm{min}$ is small.  Therefore, $z \longrightarrow \exp( -X ) \,-1 \approx \, -X$.  Figure~\ref{fig:5} is a plot of $D_\mathrm{a}$ versus $X$.  The straight line is a plot of the least squares fit of the data.  The line is 
\begin{eqnarray}
D_\mathrm{a}&=& (-2600 \pm 100 \mathrm{Mpc} ) X + (0.9 \pm 0.2 \mathrm{Mpc}) \nonumber \\*
&\approx& \frac{c}{115} z
\label{eq:28}
\end{eqnarray}
at 1$\sigma$ and with a correlation coefficient of 0.87.

At $D_\mathrm{a} =23$ Gpc, Eq.~(\ref{eq:28}) implies $\exp(X) = K_\mathrm{min} / 2$.  At large cosmological distance, $z \longrightarrow \,K_\mathrm{min}^{-1} \approx 1000$.  Therefore, the number of hods in photons from super cosmological scale distances are approximately equal and the universe \emph{appears} isotropic.

Homogeneity in the CUM means the density of Space and the number and strengths of galaxies are approximately the same in volumes with radii of order of hundreds of Mpc.  Therefore, in cosmological scale distances, the universe \emph{appears} statistically homogenous \cite{peac2}.  

However, this suggestion is based on the assumption that the more distant volumes of Space are similar to our local volume.  Since distance implies regression in time, the $M_\mathrm{g} =0$ assumption may be invalid at cosmological scale distances.  Since distance implies a larger $\Delta N / N$, the $C_\mathrm{s}$ varies at cosmological scale distances.

The values of $X$ are less than zero.  The blueshift is because the 1/[$K_\mathrm{min} - \exp(X)$] term is less than one.  Since the $\mathrm{d} \rho / \mathrm{d} t$ term is small for inter-galaxy distances, the energy continuity equation implies the existence of a physical mechanism to add hods to photons as they travel from emitters.  Further, the $\nabla \rho$ and $\nabla ^2 \rho$ may equal zero.  This suggests there may exist regions of the universe wherein the number of hods in transiting photons remain unchanged. 

The calculation of the Doppler shift herein assumes light is corpuscular rather than wave-like.

\section{\label{sec:disc}Discussion}

The correlation coefficient of 93\% was obtained with 14 constants, of which six are galaxy parameters, of Eq.~(\ref{eq:26}) for the 32 trial galaxies.  For the Hubble law to achieve such a correlation coefficient for the 32 trial galaxies, 33 parameters would be required.  In addition, the CUM explains the galaxy streaming noted by \citet{aaro2}, \citet{lilj}, and \citet{lynd}.

The requirement to know the position and absolute magnitude of all galaxies, particularly those galaxies between the target galaxy and us, makes the use of Eq.~(\ref{eq:26}) impractical for larger distances.  The first approximation at large distances (the Hubble law) is more practical unless Sinks or galaxies are near the photon path such as when gravitation lensing occurs.

Although statistical homogeneity in matter and Space distribution on large scales appears valid, on smaller scales matter and Space are inhomogeneous.  Therefore, a cosmology that explains small-scale observations (including inhomogeneity), that reduces to statistical homogeneity on cosmological scales, and that reduces to Newtonian expectation on very small scales seems more plausible than a cosmology that requires homogeneity and that fails on galaxy scales~\cite{bell,bosc,sell}.  

\section{\label{sec:conc}CONCLUSION}

The redshift - distance ($z-D$) relation derived from the Changing Universe Model (CUM) has been successfully demonstrated for a trial sample of 32 galaxies with distances calculated using Cepheid variable stars.  The equation developed is 
\begin{eqnarray}
\frac{1}{z+1}&=& K_\mathrm{min} + \exp( K_\mathrm{dp} D P + K_\mathrm{p} P + K_\mathrm{f} F \nonumber \\*
& &+ K_\nabla \int_0^D \! \nabla ^2 \rho \, \mathrm{d} x + K_\mathrm{vp} P v_\mathrm{e} )
\label{eq:40},
\end{eqnarray}
where $P$, $F$, and $\nabla ^2 \rho$ depend on the Space and gravitation forces and the $K$ values are constants.

The fact that the CUM successfully delivers a high correlation $z-D$ relation offers hope the CUM and its inhomogeneous character may extend to cosmological scale distance.

\begin{acknowledgments}
This research has made use of the NASA/IPAC Extragalactic Database (NED) which is operated by the Jet Propulsion Laboratory, California Institute of Technology, under contract with the National Aeronautics and Space Administration.

This research has made use of the LEDA database (http://leda.univ-lyon1.fr).

This research has made use of the NASA's Skyview facility located at NASA Goddard Space Flight Center.

I acknowledge and appreciate the financial support of Maynard Clark, Apollo Beach, Florida, while I was working on this project.
\end{acknowledgments}

\appendix

\section{\label{sec:initial}INITIAL DISTANCE ESTIMATION}
Category A galaxies consisted of the 32 trial galaxies. 
\begin{figure}
\includegraphics[width=0.5\textwidth]{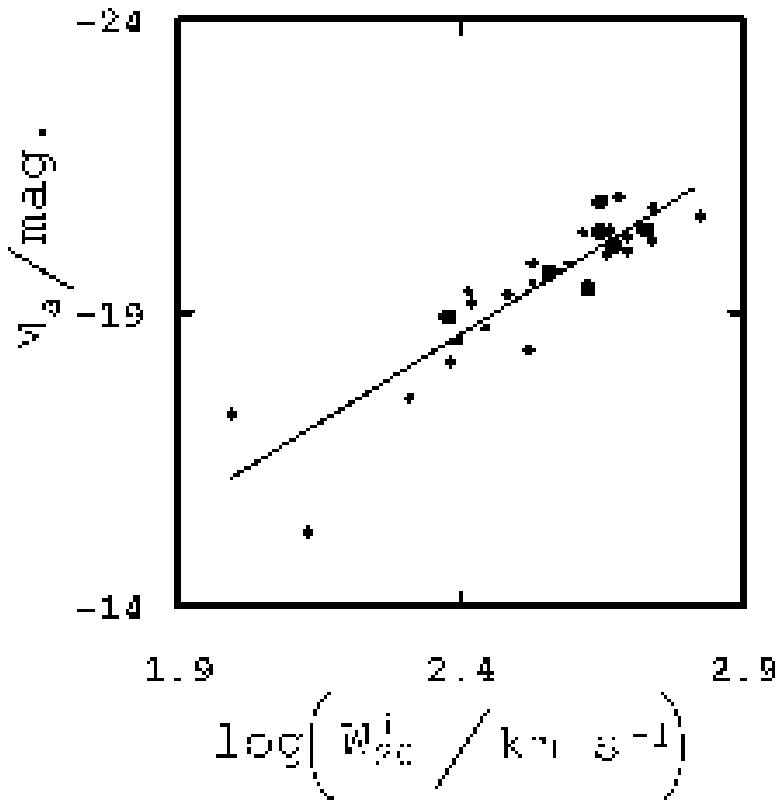}
\caption{\label{fig:2} Plot of the absolute magnitude $M_\mathrm{a}$ versus the inclination corrected 21 cm. line width $W^i_\mathrm{20}$ at 20\% of peak value for 29 of the 32 trial galaxies \cite{free,macr}.  The straight line is a plot of Eq.~(\ref{eq:31}).  The circles indicate the data points for galaxies with (l,b) = (290$^\circ \pm 20^\circ$,75$^\circ \pm 15^\circ$).}
\end{figure}
\begin{figure}
\includegraphics[width=0.5\textwidth]{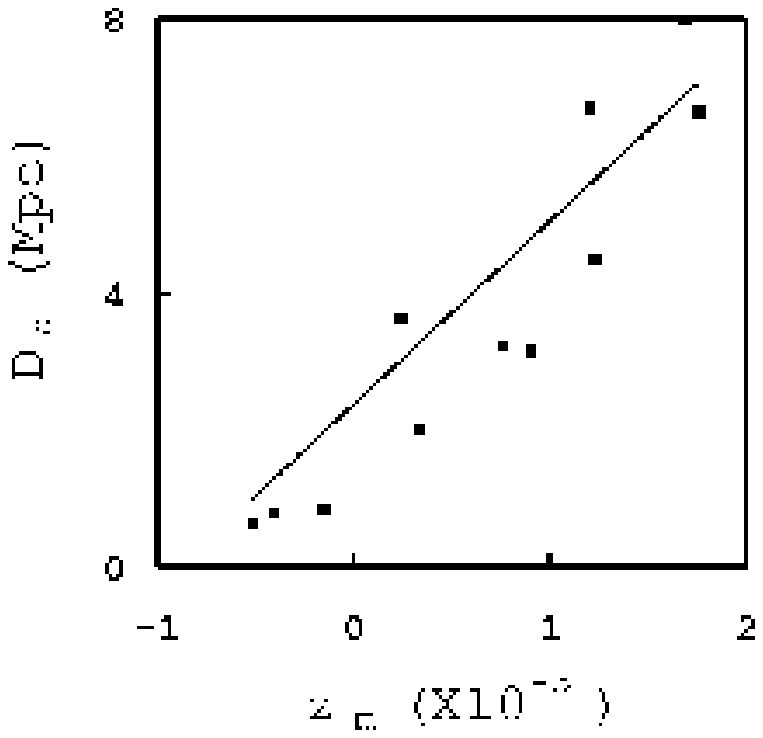}
\caption{\label{fig:3} Plot of distances $D_\mathrm{a}$ to galaxies calculated using Cepheid variable stars as listed by \citet{free} and \citet{macr} versus the redshift $ z_\mathrm{m}$ of these galaxies.  The straight line is a plot of Eq.~(\ref{eq:34}).}.
\end{figure}

Category B galaxies consisted of 5967 spiral galaxies in the sample that were not Category A galaxies with $W_\mathrm{20}$, $i_\mathrm{n}$, and $m_{\beta }$ values listed in the databases.  The distance $D_\mathrm{b}$ (Mpc) for each of the Category B galaxies were calculated as follows: (1) For the Category A galaxies, the absolute magnitude $M_\mathrm{a}$ was calculated
\begin{equation}
\frac{M_\mathrm{a}}{\mathrm{mag.}}= \frac{m_{\beta k}}{\mathrm{mag.}} - \frac{E_{\mathrm{xt}k}}{\mathrm{mag.}} - 25 - 5 \, \log \left( \frac{D_\mathrm{a}}{\mathrm{Mpc}} \right)
\label{eq:30}.
\end{equation}
(2) A plot of $M_\mathrm{a}$/mag. versus $\log(W_\mathrm{20}^i/ \mathrm{km \, s}^{-1})$, where $W^i_\mathrm{20}$ is the inclination corrected $W_\mathrm{20}$, is presented in Fig.~\ref{fig:2}.  IC 1613 (an Irregular galaxy), NGC 5253 (an Irregular galaxy), and NGC 5457 (\citet{free} noted the $D_\mathrm{a}$ was calculated differently) were omitted from the plot.  The straight line in Fig.~\ref{fig:2} is a plot of 
\begin{equation}
\frac{M_\mathrm{a}}{\mathrm{mag.}}= K_\mathrm{ws} \, \log \left( \frac{W^i_\mathrm{20}}{\mathrm{km \, s}^{-1}} \right) + K_\mathrm{wi}
\label{eq:31},
\end{equation}
where $K_\mathrm{ws}= -6.0 \pm 0.6$, and $K_\mathrm{wi}=-4 \pm 1$ at one standard deviation ( 1$\sigma$).  The correlation coefficient is -0.90.  \citet{tull77} calculated a similar relation with $K_\mathrm{ws}= -6.25$ and $K_\mathrm{wi}=-3.5$.  The circles indicate the data points for galaxies with (l,b) = (290$^\circ \pm 20^\circ$,75$^\circ \pm 15^\circ$) as in Fig.~\ref{fig:1}.  However, unlike Fig.~\ref{fig:1}, the data in Fig.~\ref{fig:2} for the outlier galaxies appears consistent with the other sample galaxies' data.
(3) The $D_\mathrm{b}$ is 
\begin{equation}
D_\mathrm{b}= 10^{0.4 (m_{\beta k} - M_\mathrm{b} - E_{\mathrm{xt}k})}
\label{eq:33},
\end{equation}
where 
\begin{equation}
\frac{M_\mathrm{b}}{\mathrm{mag.}}= K_\mathrm{ws} \, \log \left(\frac{W^i_\mathrm{20}}{\mathrm{km \, s}^{-1}} \right) + K_\mathrm{wi}
\label{eq:33a}.
\end{equation}

Category C galaxies consisted of galaxies in the sample that were not in the previous categories with  $-0.001  < z_\mathrm{m} <.002$.  A plot of $D_\mathrm{a}$ versus $ z_\mathrm{m}$ for the trial galaxies in the Category C range is presented in Fig.~\ref{fig:3}.  Since the goal of this section is to arrive at the initial distance estimation, NGC 2541 and NGC 4548 are outlier galaxies and were omitted from the plot for simplicity.  The straight line in Fig.~\ref{fig:3} is a plot of 
\begin{equation}
D_\mathrm{a} = \frac{ c z_\mathrm{m}}{ K_\mathrm{czs} } + K_\mathrm{czi}
\label{eq:34},
\end{equation}
where $K_\mathrm{czs}= 100 \pm 10 $ km~s$^{-1}$~Mpc$^{-1}$, and $K_\mathrm{czi}= 1.7 \pm 0.4$ Mpc at 1$\sigma$.  The correlation coefficient is -0.93. 

The distance $D_\mathrm{c}$ (Mpc) for Category C galaxies is 
\begin{equation}
D_\mathrm{c} = \frac{ c z_\mathrm{m}}{ K_\mathrm{czs}} + K_\mathrm{czi}
\label{eq:35}.
\end{equation}

Category D galaxies are galaxies in the sample not in the previous categories and with $ z_\mathrm{m} < -0.001 $.  The distance $D_\mathrm{d}$ (Mpc) to Category D galaxies was 1 Mpc.

Category E galaxies are all other galaxies in the sample not in the previous categories.  The distance $D_\mathrm{e}$ (Mpc) to these galaxies was calculated using the Hubble law with $H_\mathrm{o}= 70$ km~s$^{-1}$~Mpc$^{-1}$.


\end{document}